\documentclass[a5paper,twocolumn]{article}
\usepackage{graphicx}
\usepackage{amsmath}
\makeatletter
\def\@maketitle{%
  \newpage
  \null
  \begin{center}%
  \let \footnote \thanks
    {\large\textbf \@title \par}%
    \vskip 1.5em%
    \lineskip .5em%
    \begin{tabular}[t]{c}%
      \@author
    \end{tabular}\par
    \vskip 1em%
    \@date
  \end{center}%
  \par
  \vskip 1.5em}
\renewcommand\section{%
  \@startsection {section}{1}{\z@}%
  {1\baselineskip \@plus 1ex \@minus .2ex}%
  {0.1ex \@plus.2ex}%
  {\normalfont\bfseries}}
\makeatother

\begin{document}
\title{Physical Model of Nernst Element}
\author{%
Hiroaki~NAKAMURA 
Kazuaki~IKEDA$^{\ast}$ and 
Satarou~YAMAGUCHI$^{\ast\ast}$ }

\date{%
 Venture Business Laboratory, Nagoya University, 
 Furo-Cho, Chikusa-Ku, Nagoya, 464-8603, Japan, \\
 Tel.\&FAX  +81-52-789-4538, 
 E-mail:hiroaki@rouge.nifs.ac.jp,
 http://rouge.nifs.ac.jp/$\tilde{\ }$hiroaki/index.html\\
 $^{\ast}$Department of Fusion Science, 
 The Graduate University for Advanced Studies,  \\
 $^{\ast\ast}$National Institute for Fusion Science (NIFS) }

\maketitle
\thispagestyle{empty}

\begin{abstract}
Generation of electric power by the Nernst effect is a new application of a semiconductor. A key point of this proposal is to find materials 
with a high thermomagnetic figure-of-merit, which are called 
Nernst elements. In order to find candidates of the Nernst element, 
a physical model to describe its transport phenomena is needed. 
As the first model, we began with a parabolic two-band model in 
classical statistics. According to this model, we selected InSb as 
candidates of the Nernst element and measured their transport 
coefficients in magnetic fields up to 4 Tesla within a temperature 
region  from  270K to 330K. In this region, we calculated transport 
coefficients numerically by our physical model. For InSb, experimental 
data are coincident with theoretical values in strong magnetic field.
\end{abstract}

\section{Introduction}

One of the authors, S. Y., proposed~\cite{yama1} the direct electric energy 
conversion of the heat from plasma by the Nernst effect in a fusion reactor, 
where a strong magnetic field is used to confine a high temperature fusion 
plasma. He called~\cite{yama1,yama2} the element which induces the 
electric field in the presence of temperature gradient and magnetic 
field, as  Nernst element. In his papers~\cite{yama1,yama2}, 
he also estimated the figure of merit of the Nernst element 
in a semiconductor model. In his results~\cite{yama1,yama2}, 
the Nernst element has high performance in low temperature region, 
that is, 300 -- 500 K.
Before his works, the Nernst element was studied in the 1960's~\cite{Harman}. 
In those days, induction of the magnetic field had a lot of loss of  energy. 
This is the reason why the Nernst element cannot be used. 
Nowadays an improvement on superconducting magnet gives us   
higher efficiency of the induction of the strong magnetic field. 
We started  a measuring system of transport coefficients 
in the strong magnetic field to estimate efficiency of the 
Nernst element on a few years ago~\cite{ikeda}.
We need criteria to find materials with high efficiency. 
The first model is one-band model which was proposed by S. Y. 
However is model cannot explain the temperature dependence 
of the Nernst coefficient above the room temperature for 
intrinsic indium antimonide, InSb\_X~\cite{ikeda,nakamura1}.
We improved the one-band model to the two-band model.
In this paper, we measured InSb\_B which is doped Te heavier 
than InSb\_X. Near room temperature, the sample InSb\_B transits 
from the extrinsic region to the intrinsic region. 
To calculate transport coefficients of InSb\_B in a magnetic field, 
we use the two-band model, In this paper, we report the calculations 
by the two-band model. 
(In Ref. ~\cite{ikeda2}, we also measured  and calculated 
transport coefficients of Ge in a magnetic field near room temperature.)
 
\section{Theoretical calculations}

As the physical model to describe transport phenomena of 
the material in the Nernst element, we use a parabolic 
two-band model in the classical statics. We have 
the following parameters of this model;
\begin{itemize}  
	\item $m_{\rm n}$ ($m_{\rm p}$):
	  effective mass of electron (hole),  
	\item $\varepsilon_{\rm D}$ ($\varepsilon_{\rm A}$)
	 : energy level of a donor (an acceptor), 
	\item $N_{\rm D}$ ($N_{\rm A}$) :
	 concentration of donors (acceptors), 
	\item $\mu_{\rm n}$ ($\mu_{\rm p}$)
	 : mobility of an electron (a hole),
	\item $\varepsilon_{\rm G}$: energy gap, 
	 $\varepsilon_{\rm F}$: fermi energy.
\end{itemize} 

Using these parameters, we obtain concentrations of carriers as follows:
\begin{eqnarray}
n(T) &=& N_{\rm C} (T) \exp\left( 
\frac{\varepsilon_{\rm F} - \varepsilon_{\rm G} }{k T} \right), \label{eq.1} \\
p(T) &=& N_{\rm V} (T) \exp \left(
\frac{- \varepsilon_{\rm F} }{k T} \right) , \label{eq.2}
\end{eqnarray}
where $n (p)$  is the concentration of free electron (hole). 
Here $N_{\rm C}$ ($N_{\rm V}$), the effective 
density of state in the conduction (valence) band is given by 
\begin{eqnarray}
 N_{\rm C} (T) &=& 2\left(  \frac{ 2 m_{\rm n} \pi k T}{h^2}
 \right)^{\frac{3}{2}}, \label{eq.3} \\
 N_{\rm V} (T) &=& 2\left(  \frac{2 m_{\rm p} \pi k T}{h^2}
 \right)^{\frac{3}{2}}. \label{eq.4} 
 \end{eqnarray}
We also obtain the concentration of electrons (holes) 
in the donor (acceptor) level, $n_{\rm D }$ ($p_{\rm A}$) as follows:
\begin{eqnarray}
n_{\rm D } &=& N_{\rm D} \frac{1}{1+ \frac{1}{2} 
\exp\left( -\frac{\varepsilon_{\rm D} -
\varepsilon_{\rm G} +\varepsilon{\rm F} }{kT} \right)}, \label{eq.5}\\
p_{\rm A} &=& N_{\rm A} \frac{1}{1+ 2 
\exp\left( -\frac{\varepsilon_{\rm A} -
\varepsilon_{\rm F}  }{kT}\right)}. \label{eq.6}
\end{eqnarray}
We suppose the charge neutrality as
\begin{equation}
N_{\rm D} - n_{\rm D} + p(T) = 
N_{\rm A} -p_{\rm A}(T) +n(T) . \label{eq.7}
\end{equation}
Substituting the concentrations of carriers with eqs. (\ref{eq.1})-(\ref{eq.6})
in eq. (\ref{eq.7}), we obtain the following algebraic equation in value 
$x \equiv  \exp \left( \varepsilon_{\rm F} /k T\right) $ as 
\begin{eqnarray}
&s u x^4 + \left( u+N_{\rm A} s +stu\right)x^3 + 
\left( N_{\rm A} -N_{\rm D} +ut-N_{\rm V} s \right) x^2 &\nonumber\\
&-\left(N_{\rm D}t+N_{\rm V}+N_{\rm D} st \right) x -N_{\rm V} t =0,&
\label{eq.8}
\end{eqnarray}
where
\begin{eqnarray}
s&=&2\exp \left(\frac{\varepsilon_{\rm D}-\varepsilon_{\rm G}}{kT}\right),
\nonumber \\
t&=&\frac{1}{2} \exp \left(\frac{\varepsilon_{\rm A}}{kT}\right),
\nonumber \\
u&=&N_{\rm C} \exp \left(-\frac{\varepsilon_{\rm G}}{kT} \right). 
\label{eq.9}
\end{eqnarray}
Using the fermi energy which is given from eqs. (\ref{eq.8}) and 
(\ref{eq.9}) , we can solve the Boltzmann equation  
of this model in a magnetic field with a perturbation 
theory and the relaxation time approximation. 
See Ref.~\cite{yama1} for details. 
Here we define the following parameters to simplify  formulation as
\begin{equation}
\eta\equiv\frac{\varepsilon_{\rm A}}{kT},
\gamma=\frac{2m_{\rm n} kT}{\hbar^2} ,
\beta_0=\frac{\sqrt{\pi} \mu_{\rm n} B }{4 z},
\beta=\frac{\beta_0}{4} \gamma^\frac{5}{2}.
\label{eq.10}
\end{equation}
We also define the following integrals as
\begin{eqnarray}
I_i\left(\beta_0\right) &=& 4 \gamma^{-1} 
\int_0^{\infty}\frac{x^i \exp \left(\eta-x\right) }{
1+ \frac{\beta_0^2}{x} }, 
\label{eq.11}\\
J_j\left(\beta_0\right) &=& 16 \gamma^{-\frac{7}{2} } 
\int_0^{\infty}\frac{x^{j-\frac{1}{2} } \exp \left(\eta-x\right) }{
1+ \frac{\beta_0^2}{x} }. \label{eq.12}
\end{eqnarray}
Using the above eqs. (\ref{eq.10}) - (\ref{eq.12}), 
we obtain transport coefficients in a magnetic field $B,$ as follows:
\begin{eqnarray}
\sigma(B)&=&\sigma(0)\frac{I_1^2+\left(\beta J_1\right)^2}{ I_1(0) I_1},
\ \  \sigma(0)= e n \mu_{\rm n}, \label{eq.13} \\
R_{\rm H}(B) &=& \frac{3 \pi^2 }{z e n}  
\frac{nJ_1}{I_1^2 +\left( \beta J_1\right)^2},
\label{eq.14} \\
\alpha(B) &=& \frac{k}{z e } \left\{ 
\frac{I_1 I_2 + \beta^2 J_1 J_2 }{
I_1^2 + \left( \beta J_1 \right)^2 } \right\} ,
\label{eq.15} \\
\beta(B)  &\equiv& N(B) B = \frac{k\beta}{z e } \left\{
\frac{J_1 I_2 - I_1 J_2 }{ 
I_1^2 + \left(\beta J_1\right)^2} \right\}, 
\label{eq.16}
\end{eqnarray}
where is the conductivity, $R_{\rm H}$
the Hall coefficient , $\alpha$ the thermoelectric power, 
and $N$ the Nernst coefficient for electron 
($z = -1$) . For hole ($z=1$), 
we must use  $p, \eta+\varepsilon_{\rm G},$
and   $\mu_{\rm p}$ instead of $n, \mu_{\rm n} $
and $\eta.$ 
Relations between these one-band transport coefficients 
and the two-band ones are written as~\cite{putly}
\begin{equation}
\sigma = \frac{D}{ \sigma_1 
\left(1+B^2 R_{\rm H2}^2 \sigma_2^2 \right)
+ \sigma_2 \left(1+B^2 R_{\rm H1}^2 \sigma_1^2 \right)} ,
\label{eq.17}
\end{equation}
\begin{eqnarray}
&R_{\rm H} = \frac{1}{D}  \hspace{7cm}&  \nonumber \\
&  \times\left\{ R_{\rm H1} \sigma_1^2 +
 R_{\rm H2} \sigma_2^2 + B^2R_{\rm H1}
R_{\rm H2} \sigma_1^2\sigma_2^2 \left( R_{\rm H1} +
R_{\rm H2} \right) \right\},& \nonumber \\ 
& & \label{eq.18} 
\end{eqnarray}
\begin{eqnarray}
&\alpha= \frac{1}{D}  \hspace{7cm}& \nonumber \\
&  \times\left\{
     \begin{array}{l}
            \alpha_1\left\{ \sigma_1\left(\sigma_1+\sigma_2\right)
	 + \sigma_1^2\sigma_2^2R_{\rm H2}
	 \left(R_{\rm H1}+R_{\rm H2} \right)
	B^2\right\}  \nonumber \\
            \alpha_2\left\{ 
	\sigma_2 \left( \sigma_1 + \sigma_2 \right)
	+ \sigma_1^2 \sigma_2^2 R_{\rm H1}
	\left(R_{\rm H1}+R_{\rm H2} \right) B^2
	 \right\} \nonumber \\
           \left. 
	+ \sigma_1 \sigma_2 \left( N_1 - N_2 \right)
	 \left( R_{\rm H1} \sigma_1 - R_{\rm H2} \sigma_2
	 \right) B^2  \right\}
       \end{array} 
       \right\} , & \nonumber \\ & & \label{eq.19}  
\end{eqnarray}
\begin{eqnarray}
&N = \frac{1}{D}   \hspace{7cm}&\nonumber \\
& \times\left\{
     \begin{array}{l}
           N_1\left\{ \sigma_1\left(\sigma_1+\sigma_2\right)
	 + \sigma_1^2\sigma_2^2R_{\rm H2}
	 \left(R_{\rm H1}+R_{\rm H2} \right)
	B^2\right\} \nonumber \\
           N_2\left\{ 
	\sigma_2 \left( \sigma_1 + \sigma_2 \right)
	+ \sigma_1^2 \sigma_2^2 R_{\rm H1}
	\left(R_{\rm H1}+R_{\rm H2} \right) B^2
	 \right\} \nonumber \\
           \left. 
	+ \sigma_1 \sigma_2 \left( \alpha_1 - \alpha_2 \right)
	 \left( R_{\rm H1} \sigma_1 - R_{\rm H2} \sigma_2
	 \right)   \right\}
       \end{array} 
       \right\} , & \nonumber \\ & & \label{eq.20}
\end{eqnarray}
where the subscripts 1 and 2  denote the contribustion from 
conduction and balence bands, respectively. 
The parameter $D$ is described as 
\begin{equation}
D \equiv \left( \sigma_1 + \sigma_2 \right)^2 + B^2 \sigma_1^2 
\sigma_2^2 \left(R_{\rm H1}+R_{\rm H2} \right)^2 . \label{eq.20.1}
\end{equation}
By the above algorithm, 
we calculate the transport coefficients in a magnetic field. 
In this calculations, we must prepare physical quantities i.e. 
effective masses, energy levels, 
concentrations of impurities, mobilities, energy gap. 
From the previous works~\cite{nakamura2}, 
we can get the following parameters:
\begin{eqnarray}
m_{\rm n } &=& 0.0152 m_{\rm 0 }, \nonumber \\
m_{\rm p } &=& 1.1140 m_{\rm 0 }, 
\nonumber \\
\varepsilon_{\rm G} &=& 0.210 {\rm eV},  \nonumber \\
\varepsilon_{\rm D} &= & 0.0007{\rm eV}, \nonumber \\
\varepsilon_{\rm A} &=& 0.002{\rm eV}, \nonumber \\
\mu_{\rm n} &=& 38000 T^{-1.5} {\rm m/V/s} , \nonumber \\
\mu_{\rm p} &=& 1056.86 T^{-1.7} {\rm m/V/s}, \nonumber \\
N_{\rm D} &=& 2.1 \times 10^{22} {\rm m^{-3}}, \nonumber \\
N_{\rm A} &=& 0,\label{eq.21}
\end{eqnarray}
where $m_0$ is the bare electron mass.
Using eq. (\ref{eq.21}), we calculate transport coefficients.

\section{Comparison between experimental and theoretical results}
We measured transport coefficients of indium antimonide 
in a magnetic field. The sample X has the electron carrier 
concentration $n = 6.6 \times10^{20} {\rm m}^{-3}$ and mobility 
$\mu_{\rm n}=21 {\rm m^2/V/s } $ at 77K. 
The sample B has $n = 2.1
\times 10^{22} {\rm  m^{-3} } $ at 77K. 

\begin{figure}
  \begin{center}
    \includegraphics[width=8cm]{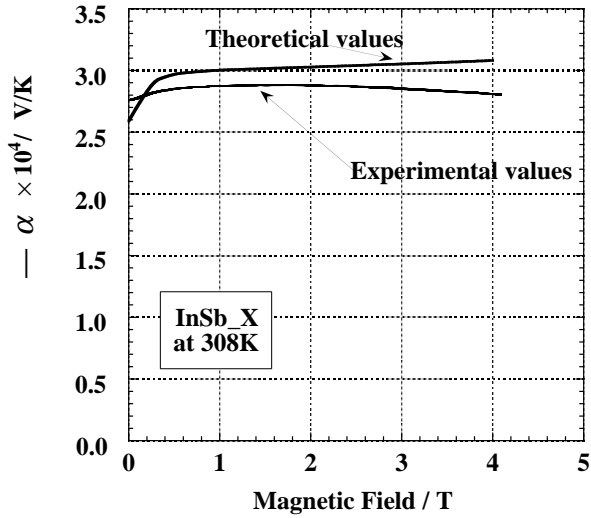}
    \caption{ Thermoelectric power versus magnetic field of InSb\_X
        at 308K }
    \label{fig.1}
  \end{center}
\end{figure}

\begin{figure}
  \begin{center}
    \includegraphics[width=8cm]{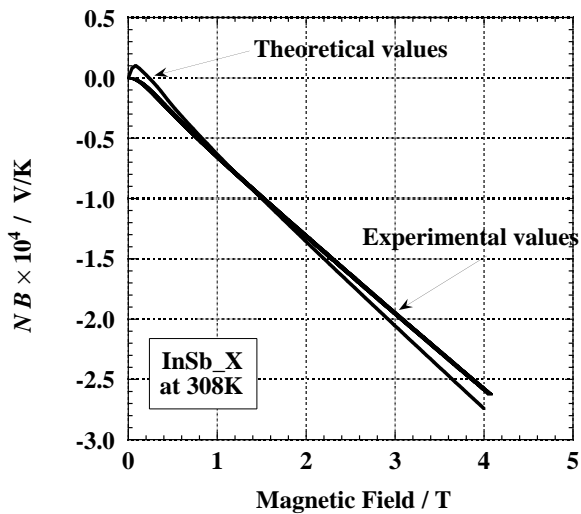}
    \caption{ Nernst Coefficient multiplied by magnetic field $NB$
    versus magnetic field of InSb\_X at 308K }
    \label{fig.2}
  \end{center}
\end{figure}

\begin{figure}
  \begin{center}
    \includegraphics[width=8cm]{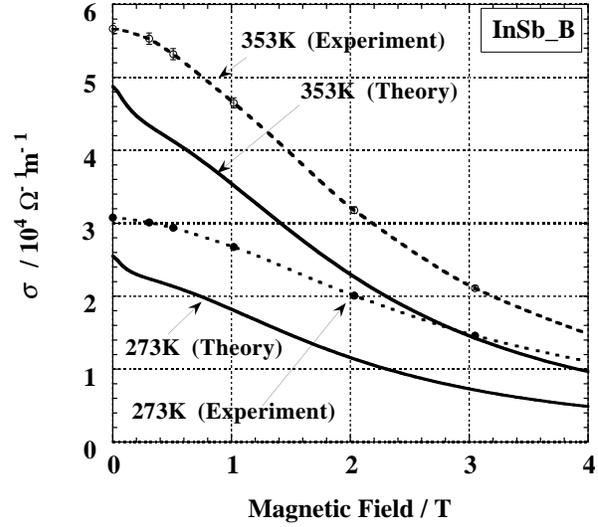}
    \caption{ Electrical conductivity versus magnetic field of InSb\_B
    at 273K and 353K }
    \label{fig.3}
  \end{center}
\end{figure}

\begin{figure}
  \begin{center}
    \includegraphics[width=8cm]{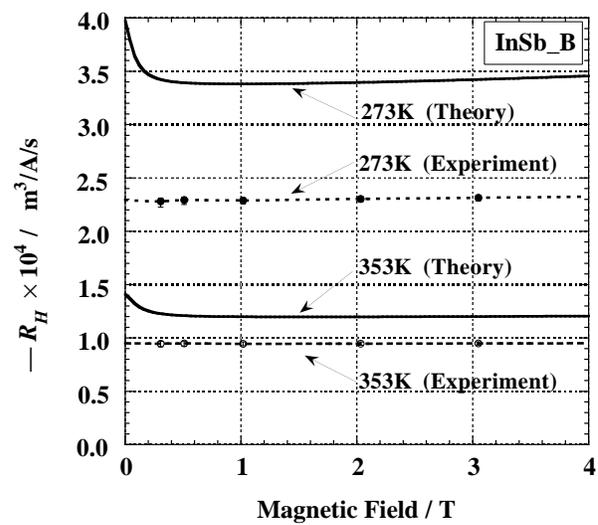}
    \caption{ Hall coefficient versus magnetic field of InSb\_B at
    273K and 353K }
    \label{fig.4}
  \end{center}
\end{figure}

\begin{figure}
  \begin{center}
    \includegraphics[width=8cm]{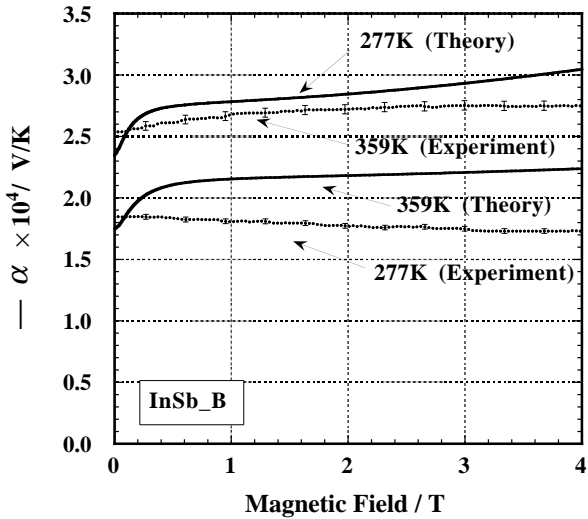}
    \caption{ Thermoelectric power versus magnetic field of InSb\_B
    at 273K and 353K }
    \label{fig.5}
  \end{center}
\end{figure}

\begin{figure}
  \begin{center}
    \includegraphics[width=8cm]{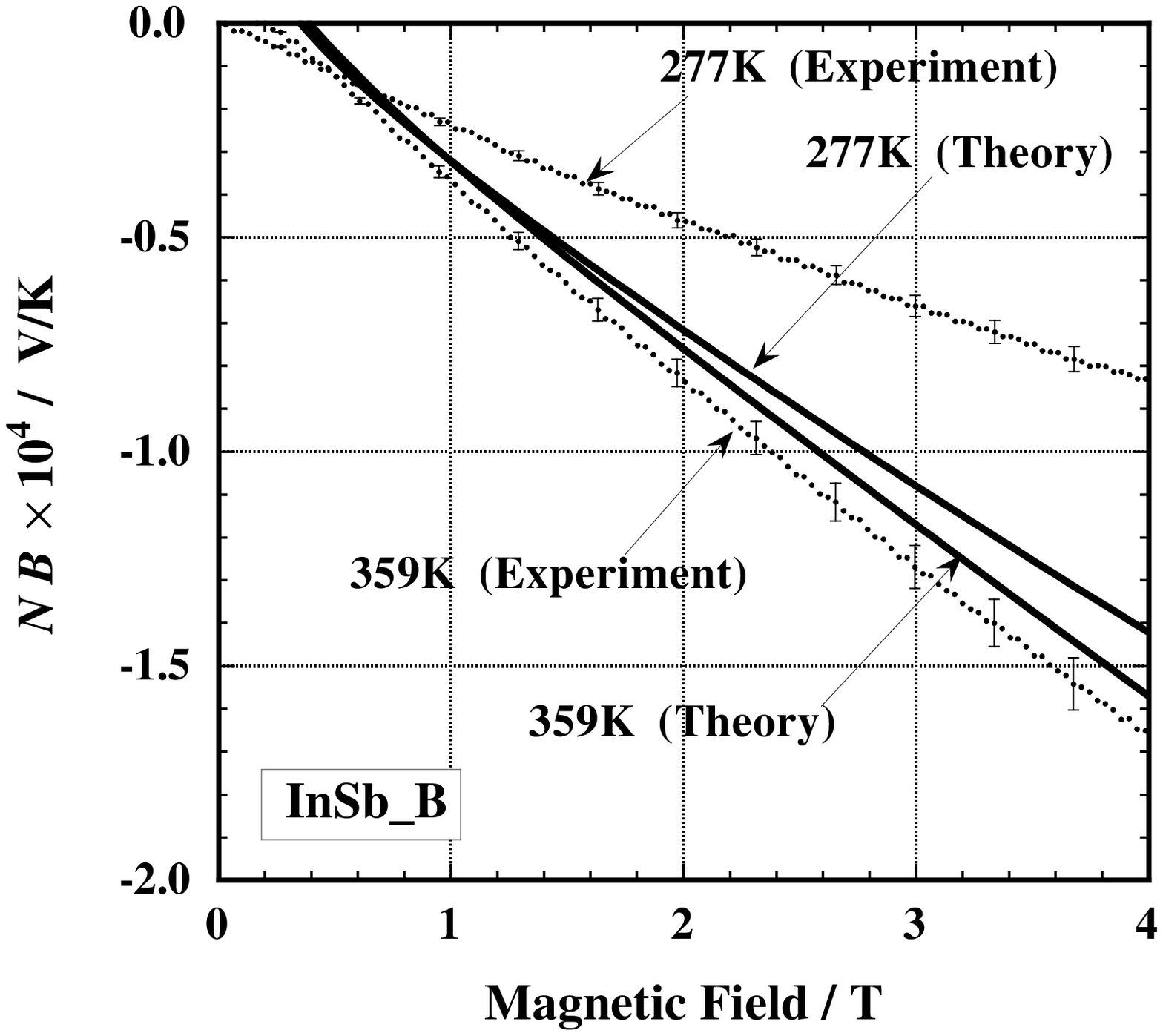}
    \caption{ Nernst coefficient multiplied by magnetic field $NB,$
    versus magnetic field of InSb\_B at 273K and 353K }
    \label{fig.6}
  \end{center}
\end{figure}

The conductivity and the Hall coefficient are measured by 
the van der Pauw method. The thermoelectric power and 
the Nernst coefficient are also measured for 
the Òbridge shapedÓ sample~\cite{nakamura2}. 
In Fig. ~\ref{fig.1}, we plot the thermoelectric power of 
InSb\_X as a function of magnetic field. 
The Nernst coefficient of InSb\_X is plotted in 
Fig. ~\ref{fig.2}. These figures show 
that these transport coefficients can be calculated 
by the two-band model. 
For InSb\_B, we also measured the conductivity, 
the Hall coefficient the thermoelectric power and 
the Nernst coefficient. These results are plotted 
in Figs.~\ref{fig.3}--\ref{fig.6}. 
These transport coefficients given by 
the theoretical calculations are coincident 
with the experimental values.

\section{Discussion and conclusions}
From comparison the experimental and the theoretical values, 
we conclude that the two-band model is enough good model 
to estimate the transport coefficient. 
We need to measure thermal conductivity 
to estimate the thermomagnetic (i.e. Nernst ) 
figure-of-merit $Z_N= \sigma (N B)^2 /  \kappa.$ 
The thermal conductivity has phonon 
scattering mechanism. We, therefore, improve 
the physical model to include the phonon 
scattering phenomena. This is a future problem.
\\
\\
\noindent
{\bf Acknowledgments}\\
The authors are grateful to Dr. Tatsumi in 
Sumitomo Electric Industries.  We appreciate 
Prof. Iiyoshi and Prof. Motojima in the 
National Institute for Fusion Science for his helpful comments.

\end{document}